\documentclass[aps,groupedaddress,nofootinbib,showkeys,showpacs,amsfonts,eqsecnum,preprint]{revtex4}

\usepackage{epsfig}
\usepackage{graphicx}
\usepackage{bm}
\usepackage{amsmath}
\usepackage{amsfonts}
\usepackage{amssymb}

\renewcommand{\Im}{{\rm Im \,}}
\newcommand{\tr}{{\rm tr \,}}
\newcommand{\Tr}{{\rm Tr}}

\newcommand{\Dirac}{{\bf D}}

\newcommand{\thru}[1]{\hbox{$ \mathrel{\mathop{#1\!\!\!\!/}}$}}
\newcommand{\thrur}[1]{\mathrel{\mathop{#1\!\!\!/}}}
\newcommand{\thrul}[1]{\mathrel{\mathop{#1\!\!\!\!\!/}}}
\newcommand{\eq}[1]{eq.~(\ref{eq:#1})}

\newcommand{\comments}[1]{} 

\begin{document} 

\title{Leading order one-loop $CP$ and $P$ violating effective action in the
  Standard Model}

\author{L. L. Salcedo}
\email{salcedo@ugr.es}

\affiliation{ Departamento de F{\'\i}sica At\'omica, Molecular y
Nuclear, Universidad de Granada, E-18071 Granada, Spain }

\date{\today}

\begin{abstract}
The fermions of the Standard Model are integrated out to obtain the effective
Lagrangian in the sector violating $P$ and $CP$ at zero temperature. We
confirm that no contributions arise for operators of dimension six or less and
show that the leading operators are of dimension eight. To assert this we
explicitly compute one such non-vanishing contribution, namely, that with
three $Z^0$, two $W^+$ and two $W^-$. Terms involving just gluons and $W$'s
are also considered, however, they turn out to vanish in the $P$-odd sector to
eighth order. The analogous gluonic term in the $CP$-odd and $P$-even
($C$-odd) sector is non-vanishing and it is also computed. The expressions
derived apply directly to Dirac massive neutrinos. All $CP$-violating results
display the infrared enhancement already found at dimension six.
\end{abstract}



\maketitle


\section{Introduction}
\label{sec:1}
A full understanding of $CP$-violation remains a challenge and for this reason
it is a fruitful field of research both in relation with the Standard Model of
particle physics and in extensions thereof
\cite{Xing:2003ez,Buras:1997fb,Neubert:1996qg,Grossman:1997pa,Winstein:1992sx,%
  Paschos:1989ur,Wolfenstein:1987pe,Donoghue:1987wu,Charles:2004jd}.
$CP$-violation enters in very different phenomena, like non vanishing of the
electric dipole momentum of elementary particles, baryogenesis
\cite{Sakharov:1967dj}, or, assuming $CPT$ invariance, the puzzling
$T$-violation. Yet, in the Standard Model, $CP$-violation is rather
elusive. There is no trace of it in the QCD sector, while in the electroweak
sector it enters through a small parameter in the CKM matrix for quarks
\cite{Kobayashi:1973fv}, and possibly also for leptons, for massive neutrinos
\cite{Maki:1962mu}. Even in the electroweak sector manifestation of $CP$
breaking requires a subtle combination, the Jarlskog determinant $\Delta$,
which requires order twelve in the quark (or leptons) masses and would vanish
if two up-like or two down-like quarks were degenerated in mass
\cite{Jarlskog:1985ht}. In any case only through fermions $CP$ can be broken
in the Standard Model. The structure of the Standard Model action implies that
integration of the fermions results in an effective Lagrangian of the form (we
assume the unitary gauge throughout)
\begin{equation}
\mathcal{L}^\text{eff}
\comments{_{CP\text{-odd}}}
(x) = \sum_\alpha g_\alpha
\left(\frac{v}{\phi(x)}\right)^{d_\alpha-4}\mathcal{O}_\alpha(x) ,
\end{equation}
where $\mathcal{O}_\alpha(x)$ represents any possible operator, of mass
dimension $d_\alpha$, constructed as a Lorentz and gauge invariant product of
the gauge fields, their derivatives and derivatives of the Higgs
field. $g_\alpha$ is the operator coupling constant, with mass dimension
$4-d_\alpha$. $\phi(x)$ denotes the Higgs field and $v$ its vacuum expectation
value. The coupling constant (which may vanish for some operators) has two
additive contributions, one from the quark loop and another from the lepton
loop. In the $CP$-odd sector, $g_\alpha$ must contain the Jarlskog
determinant. In terms of the Yukawa coupling this yields a tiny dimensionless
number, $\Delta/v^{12}$, of the order of $10^{-24}$. This fact has occasionally
been presented as an indication of an intrinsic limitation of the Standard
Model to produce enough $CP$-breaking to account for observations, including
the baryon asymmetry. While this might be true, qualitative arguments should
eventually be supported by a detailed computation.  Smit argued in
\cite{Smit:2004kh} that the coupling $g_\alpha$ is just a homogeneous function
of the quarks (or leptons) masses of the appropriate degree. This implies that
$g_\alpha \sim \Delta \times I_\alpha$, where $I_\alpha$ has a large negative
degree to compensate that of $\Delta$. Both $\Delta$ and $I_\alpha$ depend
only on the fermion masses and do not involve $v$. On the other hand, the
various quark masses are very different and widely different result can be
obtained by combining them at random. Actual calculations have been carried
out in \cite{Hernandez:2008db,GarciaRecio:2009zp} for operators of dimension
six, which is the first possible $CP$-violating contribution at one-loop.
They show that $g_\alpha \sim J\kappa/m_c^2$ where is $m_c$ the charm quark
mass, $J=2.9(2)\times 10^{-5}$ is the Jarlskog invariant
\cite{Nakamura:2010zzi} and $\kappa$ is a dimensionless coefficient of the
order of unity. Implications for cold electroweak baryogenesis have been
considered in \cite{Tranberg:2009de,Tranberg:2010af}. Unfortunately, these
two references differ in that \cite{Hernandez:2008db} finds such a dimension
six contribution in the $P$-odd sector whereas \cite{GarciaRecio:2009zp} finds
a contribution in the $C$-odd sector but none in the $P$-odd one.

The purpose of this note is manyfold. First, to reduce to the simplest and
more transparent terms the calculation of these couplings constants. Second,
to confirm that, although dimension six $CP$-odd and $P$-odd operators do
exist, their coupling vanish in the Standard Model. Third, to verify that the
order six cancellation is accidental, and non vanishing contributions in the
$CP$-odd and $P$-odd sector appear for the first time at dimension eight. The
purely gluonic leading (eighth) order term is also computed since it is
particularly simple. As it turns out, this term breaks $C$ but not
$P$. Lastly, to verify that the enhancement (as compared to the naive
estimate) found at order six is displayed also at higher orders.

\section{The method}
\label{sec:2}

We will integrate out the fermions in the Standard Model to extract the $CP$
violating contribution of the resulting effective action. This is the one-loop
approximation to the effective action with full one-particle irreducible
bosonic lines and vertices. We work at zero temperature. Quarks will be
explicitly considered. Leptons would not contribute to the $CP$-odd sector if
neutrinos are assumed to be exactly massless. For massive Dirac neutrinos the
contribution of the leptons will be completely analogous to the one obtained
for quarks.

The quark-sector Lagrangian of the Standard Model, in its Euclidean version
and in the unitary gauge, can be written as \cite{Huang:1992bk}:
\begin{equation}
{\cal L}(x) = 
\bar{q}(x)\Dirac q(x)
=
(\bar{q}_L,\bar{q}_R) \left(\begin{matrix}
 m & \thru{D}_L  \\
\thru{D}_R & m
\end{matrix}
\right)
 \left(\begin{matrix}
 q_R  \\ q_L
\end{matrix}
\right).
\end{equation}
Here $q_{L,R}$ carry Dirac, generation (family), $ud$ and color
indices ($ud$ space distinguishes the up-like from down-like quarks in each
generation). Expanding further the matrices in $ud$ space:
\begin{eqnarray}
m &=&
\left(\begin{matrix}
\frac{\phi}{v} m_u &0
\\
0 & \frac{\phi}{v} m_d
\end{matrix}
\right)
,
\quad
\thru{D}_L = 
\left(\begin{matrix}
\thru{D}_u + \thru{Z} + \thru{G} & \thrul{W}{}^+ C
\\
\thrul{W}{}^-C^{-1} & \thru{D}_d - \thru{Z} + \thru{G}
\end{matrix}
\right)
,
\quad
\thru{D}_R = 
\left(\begin{matrix}
\thru{D}_u + \thru{G} & 0
\\
0 & \thru{D}_d + \thru{G}
\end{matrix}
\right)
.
\label{eq:2.2}
\end{eqnarray}
Here $m_{u,d}$ are the diagonal matrices (in generation space) with the
up-like and down-like quarks masses, respectively. $G_\mu$ the gluon field,
$Z_\mu$ the $Z^0$ field, $W_\mu^\pm$ the $W$ boson fields, $C$ is the CKM
matrix, finally $(D_\mu)_{u,d}= \partial_\mu+q_{u,d}B_\mu$ where $q_u=2/3$,
$q_d=-1/3$, and $B_\mu$ is the weak hypercharge gauge connection. For
convenience, in all cases the coupling constant has been included in the
corresponding gauge connection. Further details can be found in
\cite{GarciaRecio:2009zp}.

After integration of the quark loop, the corresponding Euclidean effective
action is just
\begin{equation}
\Gamma = -\Tr\log\Dirac
.
\end{equation}
$\Gamma$ is the sum of the all Feynman graphs with one quark loop and any
number of bosonic legs, gauge fields and Higgs. This sum is written as a
functional which will be expressed within a covariant derivative expansion of
these bosonic fields.

Certainly, the effective action can be computed following the efficient method
outlined in \cite{GarciaRecio:2009zp} and based on \cite{Salcedo:2008tc},
applied there to sixth order in the derivative expansion. However, one of our
goals here is to present a derivation as transparent as possible, and closer
to the method introduced in \cite{Salcedo:2000hx} on which the calculation of
\cite{Hernandez:2008db} is based. To this end, we will use the relation
\begin{equation}
\delta\Gamma = 
-\Tr(\delta\Dirac\, \Dirac^{-1})
=
-\int d^4x \,\tr\!\!\left[\delta\Dirac\langle x| \Dirac^{-1}|x\rangle\right]
.
\label{eq:2.4}
\end{equation}
In the second equality it has been used that the variation $\delta\Dirac$,
induced by the variation in the gauge and Higgs fields in $\Dirac$, contains
no derivatives. The method is then to choose a suitable variation of these
fields, compute $\delta\Gamma$ in the desired sector, and subsequently seek a
functional fulfilling such a variation. The virtues of this approach are i)
the current $\langle x| \Dirac^{-1}|x\rangle$ is easier to obtain than the
$\Tr\log\Dirac$ itself, ii) the condition on $\delta\Gamma$ of being a
consistent variation provides a nontrivial check of the calculation, and iii)
even if one where to compute $\Gamma$ directly, the simplest way to avoid
integration by parts identities (i.e., redundant operators in the final
expression) is to obtain its functional derivative,
$\delta\Gamma/\delta\Dirac=-\langle x| \Dirac^{-1}|x\rangle$. This quantity is
local and so free from $x$-integration by parts identities. This issue becomes
increasingly important as the number of derivatives increases.

A convenient field to use as variation in \eq{2.4} is $Z$ which appears just
in $D^L_\mu$. We adopt such a choice, namely,
\begin{equation}
\delta\thru{D}_L=\left(\begin{matrix}
 \delta \thru{Z} & 0  \\
0 & -\delta \thru{Z}
\end{matrix}
\right)
:=\delta \thru{\hat Z},
\qquad
\delta\thru{D}_R = \delta m = 0
.
\label{eq:2.5}
\end{equation}

The Dirac operator can be written as $\Dirac = P_L \thru{D}_R P_R + P_R
\thru{D}_L P_L + P_R m P_R + P_L m P_L $ where
$P_{R,L}=\frac{1}{2}(1\pm\gamma_5)$ project on the $R$ or $L$ spaces,
respectively. $\Dirac$ can be explicitly inverted in chiral blocks. In
particular, for the $RL$ block, $P_R \Dirac^{-1} P_L = P_R (\thru{D}_L - m
\thru{D}_R\!\!^{-1} m)^{-1} P_L$. Hence, from \eq{2.4} and (\ref{eq:2.5}),
\begin{equation}
\delta\Gamma =
-\Tr (P_R \,\delta\thru{\hat Z} \,
(\thru{D}_L - m \thru{D}_R\!\!^{-1} m)^{-1} )
.
\end{equation}
Therefore, if only the $P$-odd sector (i.e., that with a $\gamma_5$) is
retained, we will have
\begin{eqnarray}
\delta\Gamma^- &=&
-\frac{1}{2}\Tr\!\left[\gamma_5 \,\delta \thru{\hat Z} \,
(\thru{D}_L- m \thru{D}_R\!\!^{-1} m)^{-1}
\right]
\nonumber \\
&=&
-\frac{1}{2}\Tr\!\left[\gamma_5 \,\delta\thru{\hat Z} 
\, (
\thru{D}_L\!\!^{-1}
+
\thru{D}_L\!\!^{-1}
m
\thru{D}_R\!\!^{-1}
m
\thru{D}_L\!\!^{-1}
+ \cdots
)\right]
.
\end{eqnarray}

It can be noted that each term of the expansion between parenthesis starts and
ends with a label $L$. Moreover, as the chiral label propagates through the
term it flips (from $L$ to $R$ and vice versa) at $m$ but not at $D_R$ or
$D_L$. Keeping these rules in mind we can simply write
\begin{eqnarray}
\delta\Gamma^- &=&
-\frac{1}{2}\Tr\!\left[\gamma_5 \,\delta\thru{\hat Z}\,
(
\thru{D}\,{}^{-1}
+
\thru{D}\,{}^{-1}
m
\thru{D}\,{}^{-1}
m
\thru{D}\,{}^{-1}
+ \cdots
)\right]
\nonumber \\
&=&
-\frac{1}{2}\Tr\!\left[\gamma_5 \,\delta\thru{\hat Z} \,
(\thru{D}+m)^{-1}
)\right]
.
\end{eqnarray}
The second equality follows from the fact that terms with an odd number of
$m$'s are automatically discarded since they cannot start and end with a label
$L$.

A simple and convenient technique to compute $\langle x| (\thru{D}+m)^{-1}
|x\rangle$ is the method of symbols
\cite{Nepomechie:1984wt,Salcedo:1996qy,GarciaRecio:2009zp}. This method is
suitable for computing one-loop Feynman graphs with external legs at zero
momentum or more generally, for expansions around zero momentum. In the
present case it takes the form
\begin{eqnarray}
\delta\Gamma^- &=&
-\frac{1}{2}\int \frac{d^4x d^4p}{(2\pi)^4} \, \tr\!\!\left[
\gamma_5 \,\delta\thru{\hat Z}\,
(
\thru{D}+i\!\thrur{p}+m)^{-1}
\right]
.
\label{eq:2.9}
\end{eqnarray}
Two remarks apply here: i) the momentum variable $p_\mu$, like $D^{R,L}_\mu$,
does not introduce a flip in the chiral label, and ii) after momentum
integration all $D_\mu$ appear only in the form $[D_\mu,~]$ and so there are
no longer differential (or pseudo differential) operators acting; only an
ordinary function of $x$ survives. At the same time gauge invariance is
ensured.

The covariant derivative expansion is just an expansion in powers of $D_\mu$.
Only even orders contribute (the space-time dimension being even). Besides, in
the $P$-odd sector the effective action starts at fourth order in four
dimensions, since a Levi-Civita pseudo tensor must be present. Hence:
\begin{eqnarray}
\delta\Gamma^- &=&
\frac{1}{2}\int \frac{d^4x d^4p}{(2\pi)^4} \, \tr\!\!\left[
\gamma_5 \,\delta\thru{\hat Z}\,
\sum_{n=0}^\infty
\left(
\frac{i\!\thrur{p}-m}{p^2+m^2}\thru{D}
\right)^{2n+3}
\frac{i\!\thrur{p}-m}{p^2+m^2}
\right]
.
\label{eq:2.11}
\end{eqnarray}

\section{Sixth order $P$-odd terms}
\label{sec:3}

Taking the appropriate values of $n$ in \eq{2.11} and the appropriate
contributions in $D^{R,L}$ and $m$, one can select the desired terms of the
effective action. At least two $W^+W^-$ pairs must be present in the $CP$-odd
sector terms, since the quark loop must visit the three generations
\cite{Kobayashi:1973fv}.\footnote{Of course, terms with a single $W^+W^-$ pair
  are allowed beyond one-loop.} A term with just $(W^+W^-)^2$ would count as
fourth order, however no $CP$-odd term can be constructed without introducing
further fields or derivatives. To sixth order such a $P$-odd and $CP$-odd term
can be written using $(W^+W^-)^2ZD$. ($D$ here refers to either $D_u$ or
$D_d$.) The question is whether this operator appears with a non vanishing
coefficient in the Standard Model or not. Ref. \cite{Hernandez:2008db} claims
that it does whereas the calculation in \cite{GarciaRecio:2009zp} concludes
that it does not. Therefore we will start by reconsidering such a term within
our present approach.

Under a variation of $Z$, the contribution of the candidate term to be found
in \eq{2.11} is of the form $\delta Z(W^+W^-)^2D$. We can set $\phi=v$, since
we are not interested in contributions from Higgs, and likewise we can set
$Z_\mu$ and $G_\mu$ to zero in $D^{R,L}_\mu$. Moreover, we can even set
$D_\mu^u=D_\mu^d=\partial_\mu$ in $D^{R,L}_\mu$. The ordinary derivatives can
be unambiguously replaced by covariant ones at the end without loss of
information since no $F_{\mu\nu}$ tensor can be present in the term
considered.

The computation is tedious but straightforward.  Let us spell out
the main steps in the calculation. We select terms with $n=1$ in \eq{2.11} and
restore the $L,R$ labels. We keep only terms starting and ending with the label
$L$ and only $m$ introduces a chiral label flip $L\leftrightarrow R$. No flip
is introduced by $m^2$, $p$ or $D$. This yields terms of the type
\begin{equation}
\delta\Gamma^- =
\int \frac{d^4x d^4p}{(2\pi)^4} \, \tr\!\!\left[
\frac{1}{2} \gamma_5 \,\delta\thru{\hat Z}\,
N \!\thrur{p} \, \thrul{\hat W}
 N m \thrur{\partial} N m \thrul{\hat W}  N \!\thrur{p} \,
\thrul{\hat W} N \! \thrur{p} \, \thrul{\hat W} N \! \thrur{p}
+\cdots
\right]
+ \text{o.t.}
\label{eq:2.12}
\end{equation}
Here we have set $\thru{D}_R={\thrur{\partial}}$ and
$\thru{D}_L=\thrur{\partial}+\thrul{\hat W}$ and $\thrul{\hat W}$ represents
the off diagonal (charged) part of $\thru{D}_L$ as a matrix in $ud$ space. We
have kept terms with four $\hat{W}$'s and one derivative. Also we have
introduced the quantity $N= (p^2+m^2)^{-1}$. The dots in \eq{2.12} refer to
further terms of the same type, while ``o.t.'' refers to other terms which
cannot have a contribution to the pattern $\delta Z(W^+W^-)^2D$.

Next, we expand the $ud$ labels using \eq{2.2} for $m$ and
$\thrul{\hat W}$, and \eq{2.5} for $\delta\hat{Z}$. This produces
\begin{equation}
\delta\Gamma^- =
\int \frac{d^4x d^4p}{(2\pi)^4} \, \tr\!\!\left[
\frac{1}{2} \gamma_5 \,\delta\thru{Z}\,
N_u \thrur{p} \,
\thrul{W}{}^+ C N_d m_d  
\thrur{\partial} N_d m_d  
\thrul{W}{}^- C^{-1} N_u \thrur{p} \,
\thrul{W}{}^+ C N_d \thrur{p} \, \thrul{W}{}^- C^{-1}
N_u  \! \thrur{p} \,
+ 
\cdots
\right]
+  \text{o.t.}
,
\end{equation}
where $ N_{u,d}= (p^2+m_{u,d}^2)^{-1} $.

At this point we can already factorize the trace between quantities which act
only in generation space, namely, $N_{u,d}$, $m_{u,d}$ and $C$, and all the
other quantities, which do not act on that space:
\begin{equation}
\delta\Gamma^- =
\int \frac{d^4x d^4p}{(2\pi)^4} \, 
\!\Big(
\frac{1}{2}
\tr[N_u C N^2_d m_d^2 C^{-1} N_u C N_d C^{-1} N_u]
\tr\!\!\left[
 \gamma_5 \,\delta\thru{Z}\,
\thrur{p} \, \thrul{W}{}^+ \! \thrur{\partial} \,
\thrul{W}{}^-  \! \thrur{p} \,
\thrul{W}{}^+  \! \thrur{p} \, \thrul{W}{}^- 
 \! \thrur{p} \,
\right]
+ 
\cdots
\Big)
+  \text{o.t.}
\label{eq:2.15}
\end{equation}

It is a general rule that $m_u$ or $m_d$ can only appear raised to even powers
and can be eliminated in favor of $N_u$ and $N_d$.\footnote{The label $u$ or
  $d$ does not change between two consecutive $W$'s. This implies two
  consecutive $D_L$ and so an even number of $m$'s. That the presence of
  $m_{u,d}$ can be obviated follows also from eq.~(7.1) of
  \cite{GarciaRecio:2009zp}.}  Eventually, all required momentum integrals and
traces on $3\times 3$ matrices in generation space can be cast in the form
\cite{GarciaRecio:2009zp}
\begin{equation}
I^k_{a,b,c,d}
=
 \int \frac{d^4p}{(2\pi)^4} (p^2)^k
\,\tr
\!\!
\left[
N_u^a C N_d^b C^{-1} N_u^c C N_d^d
C^{-1}
\right]
,
\label{eq:2.16}
\end{equation}
where the exponents $k,a,b,c,d$ are non negative integers. On the other hand,
only the $CP$-odd contribution is of interest to us. This is the component
antisymmetric under the exchange $C\to C^*$,
\begin{equation}
\hat{I}^k_{a,b,c,d}
= i\,\Im I^k_{a,b,c,d}
.
\end{equation}
Due to cyclic and hermiticity properties of the trace and matrices involved,
these integral satisfy
\begin{equation}
\hat{I}^k_{a,b,c,d}
 = 
-\hat{I}^k_{c,b,a,d}
 = 
-\hat{I}^k_{a,d,c,b}
\,.
\label{eq:2.18}
\end{equation}
Such antisymmetry under exchange of the labels $a$ and $c$, or $b$ and $d$
implies that many terms in \eq{2.15} do not have a contribution to the
$CP$-odd sector and this greatly alleviates the amount of subsequent
computation.

Performing an angular average over the momentum and taking the color and Dirac
traces in \eq{2.15} yields then,
for $CP$-odd terms ($N_c=3$ is the number of colors)
\begin{equation}
\delta\Gamma^- =
N_c\int d^4x \, 
\!\Big(
\frac{1}{3} \hat{I}^3_{1,1,2,2}
\, \epsilon_{\mu\nu\alpha\beta}
\,\delta Z_\mu
W^-_\nu
\partial_\alpha W^+_\beta W^-_\lambda W^+_\lambda
+ 
\cdots
\Big)
+  \text{o.t.}
\end{equation}

It only remains to apply the derivative on all fields at its right. It can be
checked that after this operation is carried out all terms so obtained
cancel. So there is no term of the type $(W^+W^-)^2ZD$ in the $CP$-odd,
$P$-odd sector of the Standard Model, in agreement with the alternative and
more systematic calculation in \cite{GarciaRecio:2009zp}.

\section{Dimension eight operators}
\label{sec:4}

In this section we show that at eighth order in the derivative expansion there
are non vanishing contributions in the $CP$-odd and $P$-odd sector of the
Standard Model. Concretely we consider terms of the form $(W^+W^-)^2 Z^3D$,
with no gluons nor derivatives of the Higgs field, and apply the technique
just described.  In this case we select terms with $n=2$ in \eq{2.11} and seek
terms of the type $\delta Z (W^+W^-)^2 Z^2D$. The calculation is analogous to
the one shown previously, except that now $Z$ is not set to zero in
$\thru{D}_L$, instead we use $\thru{D}_L=\thrur{\partial}+\thrul{\hat
  W}+\thru{\hat Z}$. Then we keep terms with precisely four $\hat{W}$'s, two
$\hat{Z}$'s and one derivative. After restoring the $L,R$ labels and $u,d$
labels, and carrying out the momentum integration, the trace in color, Dirac
and generation space, and applying the derivative to the right, one obtains:
\begin{equation}
\delta\Gamma^- =
N_c\int d^4x \, 
\!\Big(
2 \hat{I}^4_{1,1,2,4}
\, \epsilon_{\mu\nu\alpha\beta}
\,\delta Z_\mu Z_\lambda Z_{\nu\alpha}
W^+_\beta W^+_\lambda W^-_\sigma W^-_\sigma
+ 
\cdots
\Big)
+  \text{o.t.}
\label{eq:2.20}
\end{equation}
Here $Z_{\nu\alpha}$ stands for the $\nu$ derivative of $Z_\alpha$. We have
eliminated the $\hat{I}^3_{a,b,c,d}$ in favor of $\hat{I}^4_{a,b,c,d}$ and
have used identities involving $\delta_{\mu\nu}$ and
$\epsilon_{\mu\nu\alpha\beta}$ to bring the expression to a canonical
form. The expression contains 62 operators, each one weighted with various
integrals $\hat{I}^4_{a,b,c,d}$.

It remains to find out the effective action from which the variation in
\eq{2.20} derives. This serves also as a non trivial check of the
computation. The method is just to propose all allowed independent terms of
the form $(W^+W^-)^2Z^3D$ with arbitrary coefficients, and take a first order
variation with respect to $Z$ to fix those coefficients. The Minkowski space
result (see \cite{GarciaRecio:2009zp} for further details in the conventions)
is
\begin{eqnarray}
\mathcal{L}^{\rm eff}(x) &=&
\frac{N_c}{15}\frac{v^4}{\phi^4}
\epsilon_{\mu\nu\alpha\beta}\Big[
(12 \hat{I}_1 -16 \hat{I}_2) Z_\mu Z_\lambda^2   W^+_\nu W^+_\sigma  W^-_\alpha W^-_{\beta \sigma}
%
\nonumber\\&&
+ (4 \hat{I}_1 +23 \hat{I}_2) Z_\mu Z_\lambda^2  W^+_\sigma  W^+_{\nu \alpha} W^-_\beta W^-_\sigma 
%
+ (6 \hat{I}_1 -23 \hat{I}_2) Z_\mu Z_\lambda^2  W^+_\nu W^+_{\alpha \beta} W^-_\sigma{}^2
%
\nonumber\\&&
+(32 \hat{I}_1 +4 \hat{I}_2) Z_\mu Z_\lambda  Z_\sigma  W^+_\nu W^+_{\alpha \lambda } W^-_\beta W^-_\sigma 
%
+(16 \hat{I}_1 -38 \hat{I}_2) Z_\mu Z_\lambda  Z_\sigma  W^+_\lambda  W^+_{\nu \alpha} W^-_\beta W^-_\sigma 
%
\nonumber\\&&
+(16 \hat{I}_1 +22 \hat{I}_2) Z_\mu Z_\lambda  Z_\sigma  W^+_\lambda  W^+_\sigma  W^-_\nu W^-_{\alpha \beta}
%
+(-20 \hat{I}_1 -15 \hat{I}_2) Z_\lambda^2  Z_\sigma  W^+_\mu W^+_\sigma  W^-_\nu W^-_{\alpha \beta}
%
\nonumber\\&&
+10 \hat{I}_1 Z_\mu Z_\lambda  Z_{\nu \alpha} W^+_\sigma{}^2  W^-_\beta W^-_\lambda 
%
-20 \hat{I}_2 Z_\mu Z_\lambda  Z_{\nu \sigma} W^+_\alpha W^+_\lambda  W^-_\beta W^-_\sigma 
+\text{c.c}
\Big]
\label{eq:4.2}
\end{eqnarray}
We have defined $\hat{I}_1= \hat{I}^4_{1,1,2,4}-\hat{I}^4_{1,1,4,2}$ and
$\hat{I}_2= \hat{I}^4_{1,2,2,3}-\hat{I}^4_{2,1,3,2}$, and ``c.c'' refers to
complex conjugate; $Z_\mu$ is real, $(W^{\pm}_\mu)^*=W^{\mp}_\mu$ and
$\hat{I}_{1,2}$ are imaginary. The (underivated) Higgs field has been restored
using that it scales as the mass dimension of $\hat{I}_{1,2}$. Also the
derivative includes the field $B_\mu$ when it acts on the $W$'s
\cite{GarciaRecio:2009zp}. Numerically,\footnote{$\bar{m}_u,\bar{m}_d,m_c,m_s,m_t,m_b$
  denote the quark masses.}
\begin{equation}
\hat{I}_{1,2} = \frac{iJ}{(4\pi)^2}\frac{\kappa_{1,2}}{m_s^2 m_c^2}, \quad
\kappa_1 = 0.226,\quad \kappa_2 = 0.456.
\label{eq:2.23}
\end{equation}

We can see that the values of these coefficients are considerably larger than
simple estimates based on the Jarlskog determinant divided by the appropriate
power of $v$.  At sixth order the enhancement is driven by the small mass of
the light quarks and so this can be considered as a kind of chiral
enhancement. The possibility of such an effect was first pointed out in
\cite{Smit:2004kh} and confirmed in
\cite{Hernandez:2008db,GarciaRecio:2009zp}.

The momentum integrals $\hat{I}^k_{a,b,c,d}$ are completely explicit but
rather complicated homogeneous functions of the fermion masses
\cite{GarciaRecio:2009zp}. At sixth order these integrals are not continuous
at $\bar{m}_u,\bar{m}_d,m_s=0$, yet one can take the limit
$\bar{m}_u,\bar{m}_d\to 0$ and subsequently $m_s\to 0$ and this approximation
gives a value fairly close to the exact one \cite{GarciaRecio:2009zp}.  To
discuss the situation at eighth order we will consider the simpler case of
$m_b,m_t\to\infty$ which is a quite good approximation for $\kappa_{1,2}$.  At
eighth order the momentum integrals are more ultraviolet convergent and also
more infrared divergent than at sith order. Specifically, $\hat{I}_2$ diverges
as $1/m_s^2$ when $\bar{m}_u=\bar{m}_d=0$, with $\kappa_2=1/2$. The other
integral, $\hat{I}_1$, is more infrared divergent: in the same limit
$\kappa_1$ depends on the ratio $\bar{m}_u/\bar{m}_d$, varying continuously
between $-1/6$ for $\bar{m}_d\ll\bar{m}_u$ to $3/2$ for
$\bar{m}_u\ll\bar{m}_d$ .  In \eq{2.23} we have used $\bar{m}_u=2.55\,{\rm
  MeV}$ and $\bar{m}_d=5.04\,{\rm MeV}$.

We have also considered $CP$-violating terms containing only $W$'s and
gluons. Such terms appear for the first time at eighth order since (at one
loop) at least four $W$'s are needed to violate $CP$ and two $G_{\mu\nu}$ are
required to make a color singlet. In the $P$-odd sector one can write three
independent operators, however we find that they have zero coupling in the
Standard Model.  On the other hand, in the $P$-even ($C$-violating) sector,
there are also three operators of which one has zero coupling while the other
two terms result in the following effective Lagrangian (in Minkowski
space)\footnote{The gluon field strength tensor has been normalized according
  to $[D_\mu,D_\nu]=i(\lambda_a/2) G^a_{\mu\nu}$.}
\begin{equation}
\mathcal{L}^{\rm eff}(x)
= -\frac{4}{3} \frac{v^4}{\phi^4}\hat{I}^2_{1,1,2,2} \left(
W^+_\lambda{}^2 W^-_\mu W^-_\nu G^a_{\mu\alpha} G^a_{\nu\alpha}
- \text{c.c.}
\right)
.
\label{eq:4.4}
\end{equation}
Numerically, $\hat{I}^2_{1,1,2,2}=iJ \kappa_3/(4\pi)^2m_s^2 m_c^2$, with
$\kappa_3=3.76$. This coefficient diverges logarithmically as
$\bar{m}_u,\bar{m}_d\to 0$. Note that, at the order considered, the dimension
four gluon condensate\footnote{In the presence of the gluon condensate
  $G^a_{\mu\nu} G^a_{\alpha\beta} = \frac{1}{12} (\delta_{\mu\alpha} \delta_{\nu\beta} -
  \delta_{\mu\beta} \delta_{\nu\alpha})\langle
  (G^a_{\lambda\sigma})^2\rangle + \text{fluctuations} $.}  does not
induce a $CP$-violating interaction between the four $W$'s. Such term vanishes
identically, as it should be, since no $CP$-odd term can be written using just
$W$'s without derivatives or other fields.

\section{Conclusions}
\label{sec:6}
We have shown that, to one-loop and at zero temperature, the leading
$P$-violating $CP$-odd operators in the effective action of the Standard Model
are of dimension eight. We have computed explictly the couplings for the
operators of the form $Z^3(W^+W^-)^2$ plus one covariant derivative,
\eq{4.2}. These operators come from Feynman graphs with one quark-loop, four
$W$ legs, three $Z$ legs and possibly one $B\sim Z+\gamma$ leg. In principle,
dimension six operators could develop beyond one-loop or at finite temperature
due to the breaking of Lorentz invariance. Purely gluonic operators of
dimension eight have also been computed, \eq{4.4}, and they are $C$-odd and
$P$-even. Remarkably the corresponding coupling constants we find are not
vanishingly small, rather they have a natural scale related to intermediate
mass quarks times the Jarlskog invariant. All formulas derived for quarks
extend directly to massive Dirac leptons. This implies that even if the
neutrino masses are small their contribution to the $CP$-violating couplings
needs not be small, due to infrared sensitivity in the momentum integrals on
which the couplings depend. As a consequence, such couplings will be strongly
dependent on the mass ratios between neutrinos of the different generations.

\begin{acknowledgments}
I am grateful to C. Garcia-Recio for discussions.
Research supported by DGI under contract FIS2008-01143, Junta de
Andaluc\'{\i}a grant FQM-225, the Spanish Consolider-Ingenio 2010 Programme
CPAN contract CSD2007-00042, and the European Community-Research
Infrastructure Integrating Activity {\em Study of Strongly Interacting Matter}
(HadronPhysics2, Grant Agreement no. 227431) under the 7th Framework Programme
of EU.
\end{acknowledgments}


\end{document}